\documentclass[aps,prl,twocolumn,a4paper,showpacs,amsmath,amssymb,floatfix]{revtex4}
\usepackage{graphicx}

\newcommand{\form}{\mbox{C$_{12}$EO$_{6} \,$}}
\newcommand{\formeau}{\mbox{C$_{12}$EO$_{6}$/H$_{2}$O$\,$}}
\newcommand{\dgr}{ {\,}^{\circ} \mbox{C}}

\begin{document}

\title{Diffusion coefficients in a lamellar lyotropic phase:
evidence for defects connecting the surfactant structure}
\author{Doru Constantin}
\email{dcconsta@ens-lyon.fr}
\author{Patrick Oswald}
\affiliation{\'Ecole Normale Sup\'erieure de Lyon, Laboratoire de
Physique , 46 All\'ee d'Italie, 69364 Lyon Cedex 07, France}

\begin{abstract}

We measure diffusion coefficients in the lamellar phase of the nonionic
binary system \formeau using fluorescence recovery after photobleaching
(FRAP). The diffusion coefficient across the lamellae shows an abrupt
increase upon approaching the lamellar-isotropic phase transition. We
interpret this feature in terms of defects connecting the surfactant
structure. An estimation of the defect density and of the variation  in
defect energy close to the transition is given in terms of a simple model.

\end{abstract}





\maketitle

Topological defects are very important in condensed matter. In solids,
most of the defects are out of equilibrium and form during growth or
plastic deformation. In soft matter, like liquid crystals, defects sometimes
nucleate spontaneously in great number. In this case, they often announce
the transition towards a phase of higher symmetry.

Defective lamellar phases of lyotropic systems belong to this category.
They contain structural defects that can be point-like or linear
(dislocations). Unlike textural defects ({\em e. g.} focal conics), they are
not visible in optical microscopy but they can be investigated using
techniques such as FFEM (freeze-fracture electron microscopy)
\cite{allain1,allain2,strey1,strey2,strey3},SANS
\cite{holmes}, spin-labeling \cite{paz}, birefringence measurements
\cite{allain2,sallen1}, X-ray scattering \cite{rancon,clerc1}, NMR
\cite{schnepp}, etc. However, these methods do not give much
information about the defect topology.

For instance, in lamellar phases, three elementary point defects are
possible: ``pores'', ``necks'', and ``passages'' \cite{helfrich}. Necks
connect the non-polar medium (surfactant structure) (fig.1a), pores
(fig.1b) connect the polar medium (water), whereas passages join both
media (fig.1c). Screw dislocations are also frequent because their energy
is low \cite{kleman}. These defects also connect both media and their
core may be filled with the polar or the non-polar medium.

\begin{figure}[htbp]
\includegraphics[width=8.5cm]{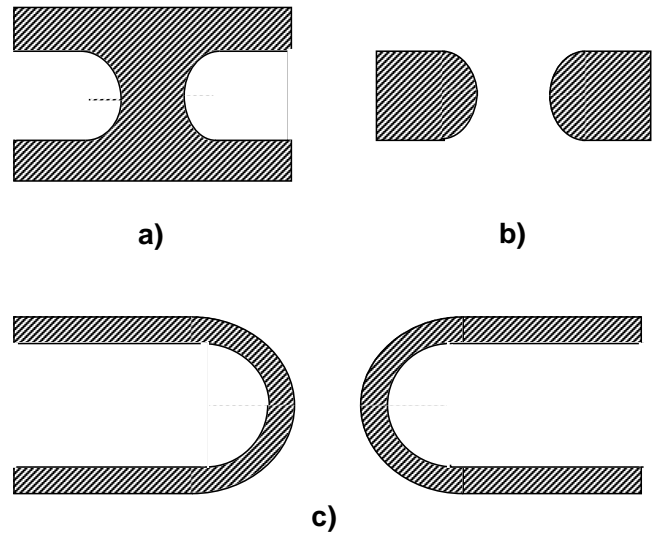}
\caption{\protect \small Elementary point defects in lamellar
phases. a) -- ``necks'', b) -- ``pores'', c) -- ``passages'' }
\label{fig1}
\end{figure}

In this Letter we present a method of determining the topology of defects
by monitoring the variation of the diffusion coefficients parallel $D_{\|}$
and perpendicularly $D_{\bot}$ to the director (normal to the layers) of
fluorescent probes that are dissolved either in the polar medium or in the
non-polar one.

The system chosen is the lamellar phase of the lyotropic mixture of the
non ionic surfactant {\form} with water. Spin-labeling measurements
\cite{paz} have shown the existence of highly curved defects, the density
of which abruptly increases a few degrees before melting. Dislocation
loops perpendicular to the layers have been observed in FFEM
\cite{allain1,allain2} ,but they can only account for a small fraction of the
total defect density. In the following we present experimental results
showing the existence of point defects. We confirm the existence of pores
\cite{paz}, and we also show that necks appear close to the lamellar --
isotropic transition.

We have first investigated the evolution of the diffusion coefficients for a
fluorescent and hydrophobic dye. If $D_{\bot}$ (parallel to the layers)
must not be significantly affected by the defects, $D_{\|}$ (across the
layers), which is very small for a perfect structure (the molecule has to
cross a water barrier), should dramatically increase if the defects connect
the surfactant structure. We performed these measures on planar domains
(director parallel to the glass plates), in order to have access
simultaneously to both $D_{\bot}$ and $D_{\|}$.

The surfactant was purchased from Nikko Ltd. and used without further
purification. We use ultrapure water from Fluka. The surfactant
concentration is {65 \%} in weight. The dye (NBD-dioctylamin) is added
in a concentration of about $0.1 \%$ and the mixture is then carefully
homogenized. The samples are prepared between two parallel glass plates,
with a spacing of 75 $\mu$m and are sealed as described in detail
elsewhere \cite{sallen1}. The lamellar phase is then oriented by use of the
directional solidification technique \cite{oswald1}. We must emphasize
that well oriented monodomains of millimetric size are needed  for our
measurements and, furthermore, we need planar anchoring on the plates.
As a general rule, the lamellar phase prefers homeotropic anchoring
(director orthogonal to the plates), but small and disorganized planar
domains can be obtained by slowly cooling the isotropic phase of the
mixture. We do that by moving the sample in the temperature gradient at
about 30 $\mu$m/s. The system is then allowed to reach equilibrium and
we continue the process at 3 $\mu$m/s. Oriented domains of the desired
size are thus obtained. We found out that the planar anchoring is very
much facilitated if the plates are ITO-covered.

Our measurement method for the diffusion constant is an adaptation of
the technique known as fluorescence recovery after photobleaching
(FRAP). The experimental setup was originally designed for the study of
thin liquid films \cite{bechhoefer}. We focus the TEM$_{00}$ mode of a
multimode Ar$^+$-ion laser (total power 70 mW) on the sample,
bleaching a spot about 40 $\mu$m in diameter.The intensity profile of the
beam is approximately Gaussian. Typical bleaching times are of the order
of 5 seconds.

The evolution of the intensity profile (proportional to the concentration
of non-bleached probe molecules $c_{\rm n} (x,y,t)$ ) is then monitored
for one minute using a cooled CCD camera, capturing 30 images. During
this time, the initial dark spot extends, due to the diffusion of bleached
molecules. The concentration profile is elliptical because of the
anisotropic diffusion. $D_{\bot}$ and $D_{\|}$ are deduced from the
images by fitting a Gaussian function. We will denote by
$c(x,y,t)=c_{\rm tot} - c_{\rm n}$ the concentration of bleached
molecules. The bleaching is considered uniform across the sample, so the
concentration obeys a two-dimensional diffusion equation (in the plane of
the sample); the $x$ axis is taken parallel to the director: $\mathbf{n} \, \|
\, {\mathbf{e}}_x$. One has :

\begin{equation}
\label{timeconc}
c(x,y,t) = C \frac{\exp \left (- \frac{x^2}{{a_1}^2 + 4D_{\|}t} -
\frac{y^2}{{a_2}^2 + 4D_{\bot}t}\right )}{\sqrt{({a_1}^2 +
4D_{\|}t)({a_2}^2 +
4D_{\bot}t)}}
\end{equation}

where $a_1$ and $a_2$ are the semi-axes of the initial bleaching spot.
The quality of the fit is crucial, since the diffusion anisotropy is very high
: $D_{\bot} \sim 100 D_{\|}$; reliable results can only be obtained with
very well oriented planar domains : small departures from the elliptical
shape (due to disorder) lead to relatively high variations in the
determined value of $D_{\|}$. We must also be sure that the NBD
molecules remain in the non-polar medium. This was shown by checking
that, in the direct hexagonal phase of the same system, NBD molecules
diffuse much faster along the cylinders than perpendicularly to them,
whereas a hydrophilic dye diffuses isotropically (water forms a
continuum among the cylinders).

\begin{figure}[htbp]
\includegraphics[width=8.5cm]{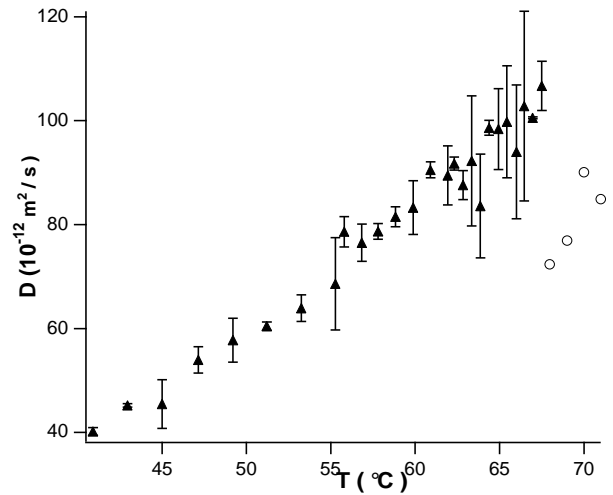}
\caption{\protect\small NBD dioctylamin. $D_{\bot}$ in the
lamellar phase (filled triangles) and $D$ in the isotropic phase
(open circles). Where present, error bars are obtained by an
average over three different measures} \label{fig2}
\end{figure}

\begin{figure}[htbp]
\includegraphics[width=8.5cm]{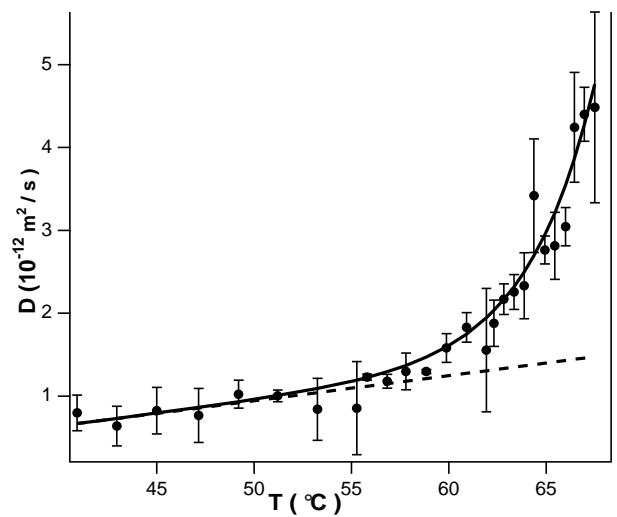}
\caption{\protect\small $D_{\|}$ of NBD dioctylamin. Axis on the
right indicates the transition temperature. Error bars are
obtained by an average over three different measures Dashed line
represents linear extrapolation of low-temperature behaviour;
solid line is exponential fit (see text).} \label{fig3}
\end{figure}

Our data in the lamellar phase are displayed in Fig. \ref{fig2} for
$D_{\bot}$ and in Fig. \ref{fig3} for $D_{\|}$. $D_{\bot}$ increases
steadily with temperature from 40 $10^{-12} \text{m} ^2 / \text{s}$ at
$41\dgr$ to 107 $10^{-12} \text{m} ^2 / \text{s}$ close to the transition
temperature ($67.7\dgr$). No pretransitional effect is observed.

The evolution of $D_{\|}$ is completely different; it remains of the order
of $10^{-12} \text{m} ^2 / \text{s}$ until about $55\dgr$, when it starts
to increase abruptly, reaching $5~10^{-12} \text{m} ^2 / \text{s}$ at the
transition. The low-temperature value could be due to probe molecules
``jumping'' the water barrier or to frozen-in defects (the density of which
does not depend on temperature), whereas the superimposed variation
($D_{\text{def}}$) clearly shows that bridges are formed between the
surfactant layers, providing a passage for the molecules. We try now to
characterize these passages as to their morphology and density.

One possibility would be that screw dislocations observed by FFEM
provide passage for the molecules, but their density at the
transition (about $100 \mu \text{m} ^{-2}$)\cite{allain2}, would
only account  for a small fraction of the measured
$D_{\text{def}}$. Indeed, a simple model for the diffusion induced
by the screw dislocations \cite{oswald2} gives :

\begin{equation}
\label{perm}
\frac{D_{\text{def}}}{D_{\bot}} \sim  \left ( \frac{r_c}{\xi}  \right ) ^2
+ 6 \pi \left (\frac{b}{\xi} \right )^2 [ 0.07 + 0.009 \ln(\xi / r_c) ]
\end{equation}

where $b = 50 {\text{\AA}}$ is the Burgers vector of the dislocations
(equal to the lamellar periodicity $\ell$ ) \cite{allain1}, $r_c$ is the core
radius (whose value is close to $b$) and $\xi = 1500{\text{\AA}} $ is the
average distance between two dislocations of opposite sign. The first term
on the r.h.s of eq.(\ref{perm}) corresponds to the ``pipe'' diffusion along
the core of the dislocations which we assume to be filled with the non-
polar medium (water cores would not contribute to diffusion). The latter
comes from the helical structure of the layers and from the fact that the
molecules can pass from one lamella to another without jumping across
the water barrier. This formula yields : $ D_{\text{def}}/{D_{\bot}}
\sim 3~10 ^{-3}$, about one order of magnitude less than the measured
value for $D_{\text{def}}/{D_{\bot}}\sim 4~10 ^{-2}$. Hence, screw
dislocations are certainly not the only type of defects present in the
sample, conclusion that confirms EPR results. Consequently, we assume
that the largest contribution to $D_{\text{def}}$ comes from point
defects of type ``neck'' close to the transition.

We now estimate the number of defects needed in order to obtain the
measured increase in $D_{\|}$. If the characteristic size of a defect is
given (roughly) by the distance between layers ($\ell \sim 50$\AA ), the
induced diffusion coefficient $D_{\text{def}}$ along the director is then
related to the defect density in the plane $n$ by :

\begin{equation}
\label{estim}
D_{\text{def}} \simeq D_{\bot}n \ell ^2
\end{equation}

within a numerical factor of order unity. Equation (\ref{estim}) simply
states that, once a molecule has diffused to the space occupied by a defect,
it crosses to the adjacent layer. Our data yield $n \simeq 0.09 \ell ^{-2} =
1200 \mu \text{m} ^{-2}$, signifying that about10 \% of the molecules
are in the defects just before the transition. This value is compatible with
the increase of the molecule proportion in the defects found by EPR
\cite{paz} within the five degrees preceding  the transition and can also
explain the birefringence drop observed in the same temperature range
and previously interpreted in terms of disorientation of the lamellae
induced by the proliferation of screw dislocation walls \cite{allain2}.

Since the defects are at thermal equilibrium, their number will be
proportional to $\exp(-E_{\text{def}} / k_B T)$, where
$E_{\text{def}}$ (the energy of the defect) decreases upon approaching
the transition. The simplest assumption is that of linear behaviour :

\begin{equation}
\label{edef}
E_{\text{def}} = \alpha (T_0 - T) + E_0 ;
\end{equation}

the fit is actually very good (figure \ref{fig3}) and gives a constant
$\alpha \simeq 100 k_B$.

\begin{figure}[htbp]
\includegraphics[width=8.5cm]{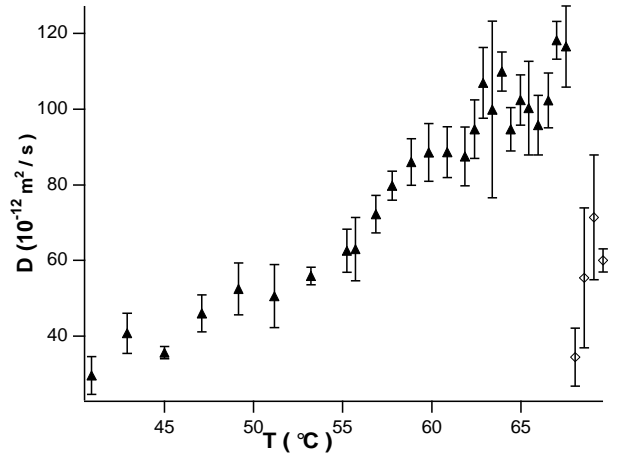}
\caption{\protect\small Fluorescein. $D_{\bot}$ in the lamellar
phase (filled triangles) and $D$ in the isotropic phase (open
circles). Error bars are obtained by an average over three to five
different measures} \label{fig4}
\end{figure}

We have also performed a complementary measure, determining the
diffusion coefficients of a hydrophilic dye (fluorescein) in the
same temperature range. The fluorescein is less bleached by the
laser beam, so the dark spot is much less contrasted than for the
dioctylamin. A shorter measure time (20 s) has been used, and the
data is more noisy; the general behaviour is, however, very clear
: while $D_{\bot}$ has roughly the same value as that in the
surfactant structure, $D_{\|}$ is roughly constant when raising
the temperature and shows no particular pre- transitional
increase. This observation definitely rules out screw
dislocations, as well as passages (fig. 1c), that connect both
media (in which case we should observe the same type of
pretransitionnal effect with both dyes).

\begin{figure}[htbp]
\includegraphics[width=8.5cm]{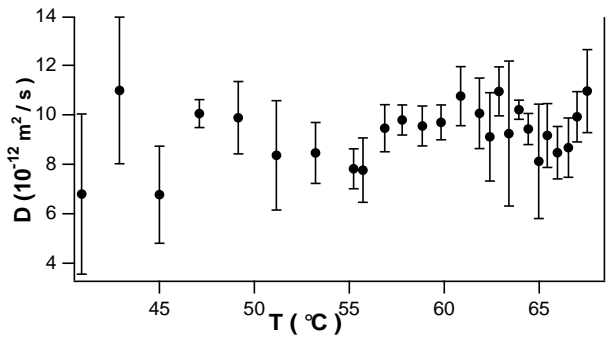}
\caption{\protect\small $D_{\|}$ of fluorescein. Axis on the right
indicates the transition temperature. Error bars are obtained by
an average over three to five different measures} \label{fig5}
\end{figure}

However, the ``necks'' are not the only defects present in the sample.
Indeed, fluorescein diffuses about 10 times faster than the dioctylamine
across the structure, whereas they have practically the same $D_{\bot}$
at all temperatures. This suggests the existence of pores facilitating
diffusion in the aqueous medium. Furthermore, the same argument that
led to eq. (\ref{estim}) gives a pore density proportional to the ratio
$D_{\|} / D_{\bot}$ for the hydrophilic dye, which would indicate that
the number of pores decreases with temperature. This interpretation is
coherent with EPR results \cite{paz} showing that the interfacial film
suffers curvature inversion a few degrees before the transition, feature
that can be interpreted as a sign of necks outnumbering pores.

One issue we have not tackled in this Letter and that will be the subject of
future work is a microscopic model for the defect energy (eq.
\ref{edef}). Theoretical treatments \cite{golubovic,gompper2} have
mainly focused upon dilute lamellar phases (while our system is, on the
contrary, very concentrated) and upon a passage-like structure for the
defect. While the passage is a defect of the bilayer, which has no
spontaneous curvature, for a neck (or a pore), one has to consider the
energy of the {\em monolayer}, the spontaneous curvature of which plays
a very important role : positive and negative values lower the elastic
energies of pores and necks, respectively \cite{helfrich}. Our results
(necks appear at higher temperature than pores) are coherent with the
general trend of non-ionic surfactants that the curvature decreases upon
heating \cite{israelachvili}. This effect has been recently measured for a
variety of water/surfactant/alkane systems \cite{sottmann}.

At this point we can also mention that necks or pores have also been
encountered in numerical simulations of lamellar phases of ternary
systems close to the transition to the microemulsion phase
\cite{gompper1,holyst,biben}.

In conclusion, we have presented unambiguous experimental evidence that
the non-polar medium in the lamellar phase of the lyotropic system
\formeau increases its connectivity when approaching the transition
towards the isotropic phase, whereas the connectivity of the water (polar
medium) decreases.  These defects are certainly of the type ``pores" at
low temperature and of type ``necks" at high temperature. These defects
add to the screw dislocations already observed via FFEM, a conclusion
that has already been drawn \cite{allain2}. The new result is that our
method allows us to distinguish between the different defect topologies,
information which is difficult to obtain with other techniques.

Though the measures have been performed on one specific lyotropic
mixture, we are confident that the FRAP measure of the diffusion
coefficients can readily be applied to other similar systems to detect
defects and their topological changes.  We are currently using this method
to characterize the defects appearing in the hexagonal phase of the same
system close to the transition to the isotropic phase.

We acknowledge fruitful discussions with Thierry Biben.

\end{document}